\pdfoutput=1
\documentclass[11pt,aps,pre,nofootinbib,nobibnotes,longbibliography,floatfix]{revtex4-2}

\usepackage{microtype}
\usepackage{graphicx}
\usepackage{amsmath}
\usepackage{amssymb}
\usepackage{xcolor}
\definecolor{myblue}{RGB}{39, 82, 180}
\usepackage[colorlinks,linkcolor=myblue,urlcolor=myblue,citecolor=myblue]{hyperref}
\usepackage{float}
\usepackage{silence}
\usepackage{amsthm}

\usepackage[ruled,vlined,linesnumbered]{algorithm2e}
\SetKwInput{KwInput}{input}
\SetKwInput{KwParam}{parameters}
\SetKw{KwRet}{return}
\DontPrintSemicolon

\newcommand{\ldstop}{LDSTOP\xspace}
\newcommand{\reals}{\mathbb{R}}
\newcommand{\pmP}{\mathbb{P}}
\newcommand{\pmQ}{\tilde{\mathbb{P}}}

\newcommand{\pX}{\tilde X}
\newcommand{\pa}{\tilde a}
\newcommand{\pp}{\tilde p}
\newcommand{\pP}{\tilde P}
\newcommand{\pJ}{\tilde J}
\newcommand{\pF}{\tilde F}
\newcommand{\pPi}{\tilde\Pi}
\newcommand{\pW}{\tilde W}
\newcommand{\pnu}{\tilde\nu}
\newcommand{\cX}{\mathcal{X}}

\DeclareMathOperator{\Var}{Var}
\newcommand{\limtoinf}[1]{\lim_{#1 \to \infty}}

\usepackage{xparse}
\NewDocumentCommand{\expec}{o m}{
  \mathbb{E}_{\IfValueTF{#1}{#1}{}}\mathopen{}\left[#2\right]
}

\begin{document}

\title{Stochastic optimisation method for estimating large deviations}

\author{Dani\"el W.\ H.\ Cloete}
\author{Hugo Touchette}
\email{htouchette@sun.ac.za}
\affiliation{Department of Mathematical Sciences, Stellenbosch University, Stellenbosch 7600, South Africa}

\newcommand{\paperdate}{November 23, 2026}
\date{\paperdate}

\begin{abstract}
A stochastic optimisation method was recently proposed to efficiently compute large deviation functions, used in statistical physics to characterise the fluctuations of nonequilibrium systems, modelled as Markov processes. The method, called \ldstop, combines numerical techniques from control theory and machine learning to iteratively construct the ``driven process'', a controlled version of the Markov process used as a model of nonequilibrium system that realises a given fluctuation or large deviation of that system in an optimal way. Compared to other methods based on spectral approximations, importance sampling, cloning or splitting, \ldstop is simple, flexible, and scalable, as it works by simulating single trajectories that are gradually guided towards the driven process. Here, we illustrate these advantages by applying the method on the full range of Markov processes considered in applications, namely, discrete-time Markov chains, continuous-time jump processes, and diffusion processes. For each type, we define the objective function to be optimised, explain different options available for representing the driven process (using, e.g., neural networks), and provide implementation details through simple applications. With these contributions, we aim to showcase the method's efficiency, as well as its ease of use when combined with available machine learning packages, such as PyTorch, TensorFlow and JAX, for simulating stochastic processes and solving high-dimensional optimisation problems.
\end{abstract}

\maketitle

\section{Introduction}

Large deviation theory provides a useful framework for studying nonequilibrium systems \cite{bertini2007,derrida2007,touchette2009,touchette2017,burenev2025}, which naturally follows from its initial development and use in the study of equilibrium systems \cite{lanford1973,ellis1985,oono1989}. In both cases, the distribution of physical quantities, such as the energy or the work done by external forces on a time-dependent system, is observed to decay exponentially with some parameter, which in the case of nonequilibrium systems is taken to be the volume (or particle number) and the observation time \cite{touchette2009,touchette2017,burenev2025}. The function that controls this decay is called the \textit{rate function} and is related to another important function in the theory, called the \textit{scaled cumulant generating function} (SCGF). Both describe the system's fluctuations as well as its steady state, so their knowledge is crucial for understanding many physical properties, including the existence of dynamical phase transitions \cite{garrahan2007,hedges2009,garrahan2009,espigares2013,tsobgni2016b}, transport coefficients \cite{gao2017,fodor2020,limmer2021}, in addition to fluctuation symmetries \cite{kurchan1998,lebowitz1999,harris2007} and dissipation bounds \cite{barato2015b,pietzonka2015,gingrich2016,gingrich2017,li2019} generalising the second law to nonequilibrium systems.

The rate function and SCGF play a role similar to the entropy and free energy and, like these thermodynamic functions, are difficult to calculate analytically \cite{touchette2009}. For this reason, many methods have recently been developed to calculate them numerically using various approaches, which can be described as being numerical or simulation-based and which involve techniques related to spectral approximations \cite{banuls2019,buca2019,casert2021,causer2021,das2021b,ferre2018,coghi2023}, optimisation \cite{ray2018,das2019,jacobson2019,oakes2020,causer2023,singh2025}, population dynamics (cloning or splitting) \cite{grassberger2002,giardina2006,cerou2007,lecomte2007a,angeli2019,brehier2019}, importance sampling \cite{kundu2011,nemoto2016,ray2018b,klymko2018,ray2020,bucklew2004,juneja2006,guyader2020}, and reinforcement learning \cite{whitelam2020,rose2021,das2021,gillman2024,pamulaparthy2025}. Among these, spectral approximations give the most accurate results, but are limited to relatively small system sizes and to discrete-state systems in one and two dimensions. For large systems of interacting particles, cloning has been mostly used, although it has severe drawbacks in terms of memory requirements, stability, and slowing-down effects appearing near phase transitions \cite{nemoto2016}. The more recent methods based on importance sampling, optimisation and machine learning prove to be more reliable and efficient, as they require the simulation of independent trajectories of a nonequilibrium system, modelled as a Markov process, which is gradually guided towards ``rare'' or ``large deviation'' regions of interest.

The method that we present in this paper is of this type: it combines importance sampling with optimisation to guide a Markov process towards rare regions of interest in an optimal way, using, as a guiding or control mechanism, the minimisation of an objective or loss function, estimated from simulated trajectories, which is also used for estimating the rate function and SCGF.  In this sense, the method can be considered as an adaptive importance sampling method guided by stochastic optimisation \cite{kappen2016}. In what follows, we refer to it as the \textit{large deviation stochastic optimisation} (\ldstop) method. 

This method was originally proposed in \cite{yan2022}, based on the theory developed in \cite{chetrite2013,chetrite2014,chetrite2015}, and was tested on two examples of diffusion processes, showing that it is more efficient than cloning, and that it can be applied at scale to processes involving many interacting particles, including active particles. Our goal in this paper is to extend this range of applications by applying \ldstop to other types of Markov processes, in particular to discrete-time Markov chains and continuous-time jump processes, used in statistical physics to model molecular motors, chemical networks, and population dynamics, among many other applications \cite{gardiner1985,amir2021,limmer2024}. For each type of process, we give the form of the objective function to minimise, describe ways of representing the guided or controlled process using various function representations, including neural networks, and test the method on simple examples to show how it works and performs in practice \cite{cloete2025}. 

From our experience with numerical methods, we find that \ldstop has many advantages. First, and most importantly, it is efficient, since it is based on single and independent trajectory simulations.
Second, it is simple and can be coded easily, especially with available packages in machine learning, such as PyTorch, TensorFlow and JAX, which provide useful interfaces for integrating neural networks in stochastic simulations, and for optimising high-dimensional objective functions using automatic differentiation or adjoint methods for computing gradients. We use these packages in our implementation of \ldstop, available on GitHub.\footnote{\href{https://github.com/DanielWHCloete/ldstop}{github.com/DanielWHCloete/ldstop}\label{github}} Finally, the method is generally stable and refinable, as it is based on solving an optimisation problem. By refinable, we mean that a given simulation can be improved upon, in principle, by trying another simulation with a different function representation or set of parameters for the controlled process, and by seeing if a lower minimum is achieved. 

We demonstrate these points in Sec.~\ref{secapps} using simple applications to illustrate the method in the most direct and pedagogical way. Before we get to this point, we recall in the next section the definitions of the rate function and SCGF for the specific large deviation problem that we consider, involving the long-time or stationary limit of additive functionals of Markov processes, which represent, physically, quantities such as the work or entropy production measured or observed relative to some nonequilibrium system. With this background, we present the \ldstop method or algorithm in Sec.~\ref{secstochopt}, providing more details on the definitions of the required estimators and available methods of computing gradients from simulations. We illustrate and test the algorithm in Sec.~\ref{secapps}, and then conclude in Sec.~\ref{secconc} with some remarks on the efficiency of \ldstop as compared with other methods, possible tests for checking its optimality, and finally, how to include error bars.

\section{Large deviation problem}
\label{secldt}

We review in this section the variational representation of the rate function and SCGF that forms the basis of \ldstop \cite{chetrite2015}. We first define the class of Markov processes and integrated observables that we consider, and then define the large deviation problem underlying these functions, which is concerned with the scaling of the distribution of observables in the long-time limit (Donsker--Varadhan-type large deviations).\footnote{The long-time regime of large deviations is different from the low-noise regime of large deviations, associated with the problem of finding transition pathways in equilibrium and nonequilibrium systems (Freidlin--Wentzell-type large deviations) \cite{touchette2009}.} For the purpose of the presentation, we take the process to be a Langevin-type diffusion process and define the class of observables specifically in that context. In Sec.~\ref{secapps}, we explain how the notations and theory change, as we apply \ldstop to other types of Markov processes.

\subsection{Markov process and observable}

We consider a Markov process $(X_t)_{t\geq 0}$ evolving in $\reals^n$ according to the following stochastic differential equation (SDE):
\begin{equation}
    dX_t = F(X_t)\,dt + \sigma\,dW_t,
\end{equation}
where $F:\reals^n \to \reals^n$ is a vector field representing the force or drift driving the system in the absence of noise, $W_t$ is a Brownian motion in $\reals^m$ modelling the environmental noise, and $\sigma$ is an $n\times m$ matrix, which we take to be constant, for simplicity, so the noise is treated as being additive.\footnote{The case of multiplicative noise, with $\sigma$ depending on $X_t$, can also be treated within large deviation theory; see \cite{chetrite2014}.} This SDE is used in physics to model various types of nonequilibrium systems involving one or many particles driven by external reservoirs and forces. It is also used in many other areas, including engineering and finance, to model systems perturbed by noise that evolve diffusively. The precise system modelled determines the form of $F$ and $\sigma$. 

For such a system, we are interested in studying the fluctuations of an \textit{observable}, defined as an integral of its state $X_t$ having the general form
\begin{equation}
    A_T = \frac{1}{T}\int_0^T f(X_t)\,dt + \frac{1}{T}\int_0^T g(X_t)\circ dX_t.
    \label{eqobs1}
\end{equation}
Here, $f$ and $g$ are two functions that define the precise observable being measured over a time $T$. As an example, in cases where $F$ is interpreted as a force, the mechanical work done per unit time (i.e., the power expended) is defined by choosing $f=0$ and $g=F$. 

Other types of observables are defined similarly by choosing $f$ and $g$ according to the system and observable considered.  In general, we can restrict ourselves to real observables, so that $f:\reals^n \to \reals$ is a scalar field, whereas $g:\reals^n \to \reals^n$ is a vector field, multiplied by the increment $dX_t$ using a scalar product that is interpreted in Stratonovich convention, as indicated by the circle symbol $\circ$ above.\footnote{This convention can be changed to another stochastic convention if needed; the Stratonovich convention is convenient because it leads to a simple expression for the long-time limit of $A_T$, involving the stationary distribution and current of $X_t$. This is explained in the next section.} This defines the class of observables normally considered for SDEs.
For other types of Markov processes, observables are defined similarly in terms of functions involving the state and jumps or transitions between states; see Sec.~\ref{secapps} for specific examples.

\subsection{Large deviation functions}

The theory of large deviations predicts, under fairly general conditions, that the distribution of $A_T$ scales as $T\to\infty$ as follows:
\begin{equation}
    \label{eqldq1}
    P(A_T=a) \approx e^{-TI(a)}.
\end{equation}
This means that the limit
\begin{equation}
    \label{eqratelim1}
    I(a) = \limtoinf{T} - \frac{1}{T} \ln P(A_T = a)
\end{equation}
exists, so the dominant term of $P(A_T=a)$ is a decaying exponential with the observation time. This approximation is useful, as the exact form of the distribution cannot be obtained for most systems of interest.

The rate of decay $I(a)$ is the rate function that we seek to calculate. In general, this function is positive and, for ergodic processes, has a unique zero $a^*$ located at the stationary expectation of $A_T$, given by
\begin{equation}
    \label{eqstata1}
    a^* = \int_{\reals^n} p^*(x)f(x)\,dx + \int_{\reals^n} J^*(x)\cdot g(x)\,dx,
\end{equation}
where $p^*(x)$ is the stationary distribution of $X_t$ solving the stationary Fokker--Planck equation, and
\begin{equation}
    J^*(x) = F(x)p^*(x) - \frac{D}{2}\nabla p^*(x)
\end{equation}
is the associated stationary current, defined with the covariance matrix $D=\sigma\sigma^\top$, with $\top$ denoting the transpose. For processes other than diffusions, the expression of $a^*$ is different, but also involves the stationary distribution and stationary current or flow; see Sec.~\ref{secapps} for examples.

The knowledge of $a^*$ is important physically as it determines the \textit{typical value} of $A_T$ around which $P(A_T=a)$ concentrates in the limit of large integration times. The rate function describes the likelihood of observing fluctuations around this value, and thus provides a complete description of these fluctuations on the large deviation (exponential) scale, which generally includes Gaussian fluctuations around $a^*$ and non-Gaussian fluctuations far from $a^*$. The rate function also gives information, as mentioned in the introduction, about the transport properties of nonequilibrium systems \cite{gao2017,fodor2020,limmer2021}, phase transitions appearing in the  fluctuations of such systems \cite{garrahan2007,hedges2009,garrahan2009,espigares2013,tsobgni2016b}, as well as general symmetries constraining the fluctuations of certain observables related to the current \cite{kurchan1998,lebowitz1999,harris2007}.

Many methods exist in large deviation theory for calculating $I(a)$. In practice, the most commonly used proceeds by calculating the SCGF, defined by the limit
\begin{equation}
    \lambda(k)= \limtoinf{T}\frac{1}{T}\ln \expec{e^{TkA_T}},
\end{equation}
where $k \in \reals$.
It is known from the G\"artner--Ellis theorem that if $\lambda(k)$ exists and is differentiable near $k=0$, then $I(a)$ exists and is given by the \textit{Legendre--Fenchel} (LF) transform of $\lambda(k)$, that is,
\begin{equation}
    \label{eqlf1}
    I(a) = \sup_{k \in \reals} \left\{k a-\lambda(k)\right\}.
\end{equation}
When $\lambda(k)$ is strictly convex, this transform reduces to a Legendre transform, given by
\begin{equation}
    I(a) = k(a)a -\lambda(k(a)),
\end{equation}
where $k(a)$ is the root of $\lambda'(k)=a$. By duality of this transform, we also have $I'(a) = k(a)$. As a result, $\lambda'(0)=a^*$. The second derivative at $k=0$ is also important as it gives the asymptotic variance of $A_T$:
\begin{equation}
    \lambda''(0) = \limtoinf{T} T\,\Var(A_T).
\end{equation}

For Markov processes, the SCGF is known to be given by the dominant eigenvalue of a matrix or linear operator, called the tilted generator, associated with the process and observable considered \cite{touchette2009,touchette2017,burenev2025}. This spectral connection, which follows from the Feynman--Kac formula \cite{touchette2017}, provides a useful numerical method for computing $\lambda(k)$, which works well in practice for low-dimensional systems. Spectral calculations can also be carried out for many-particle systems using special ansatzes, such as matrix product states or tensor networks, provided that they evolve on lattices in one or two dimensions \cite{banuls2019,buca2019,casert2021,causer2021}.  For diffusions, $\lambda(k)$ can be obtained in a spectral way essentially only in one or two dimensions \cite{mehl2008,tsobgni2016,tsobgni2018}. For this type of processes, the spectral problem is complicated by the fact that the tilted generator is generally not self-adjoint \cite{touchette2017}.

\subsection{Driven process}

The rate function cannot be estimated via direct simulations, since $P(A_T=a)$ is exponentially small with $T$, necessitating an exponentially large number of trajectories to reconstruct this distribution away from $a^*$. To avoid this difficulty, we can resort to importance sampling: that is, we can change the process being simulated in such a manner as to enhance the probability that a certain fluctuation, say $A_T=a$, arises. By reweighting techniques, we can then estimate the true probability of this event by simulating a different process $\pX_t$ \cite{bucklew2004,juneja2006,guyader2020}.

In principle, many processes can be used to transform what is a rare event for $X_t$ into a likely or typical event for $\pX_t$. Under some conditions, it is known that there exists a unique $\pX_t$ that is ``optimal'' for the purpose of estimating the rate function and SCGF using importance sampling (see \cite{guyader2020} and references therein). Finding this process, which is referred to as the \textit{fluctuation} or \textit{driven process}, is a difficult task, however, since it involves the knowledge of the spectral elements related to $\lambda(k)$ \cite{chetrite2014} or, equivalently, the solution of an optimal stochastic control problem \cite{chetrite2015}.

The latter problem is the basis of \ldstop and is defined as follows \cite{chetrite2015}. Let $\pmP_T[X]$ be the path measure associated with the trajectories or paths of $X_t$ over the time interval $[0,T]$, and let $\pmQ_T[X]$ be the similar measure describing the paths of a transformed or perturbed process $\pX_t$. In terms of these two measures, we define the relative entropy or Kullback--Leibler (KL) divergence of $\pmQ_T$ relative to $\pmP_T$ by the path integral
\begin{equation}
    D(\pmQ_T \Vert \pmP_T) =\int d[x]\, \pmQ_T[x] \ln \frac{\pmQ_T[x]}{\pmP_T[x]}
\end{equation}
over all paths evolving over $[0,T]$. The driven process is the unique process that minimises this divergence in the limit $T \to \infty$ subject to the constraint that $\expec[\pmQ_T]{A_T} = a$ \cite{chetrite2014}. In fact, in this limit, we have
\begin{equation}
    \label{eqvar1}
    I(a) = \limtoinf{T} \min_{\pmQ_T: \expec[\pmQ_T]{A_T} = a} \frac{1}{T}D(\pmQ_T \Vert \pmP_T).
\end{equation}
From this point of view, the driven process is interpreted as the controlled process $\pX_t$ closest to $X_t$ that achieves $A_T=a$ as a stationary average, and thus as a typical value of $A_T$. 
By considering the Lagrange or dual version of this constrained problem, we obtain the SCGF
\begin{equation}
    \label{eqvar2}
    \lambda(k) = \limtoinf{T} \max_{\pmQ_T} \left\{k\expec[\pmQ_T]{A_T} - \frac{1}{T}D(\pmQ_T \Vert \pmP_T)\right\}.
\end{equation}
In this result, $k$ is the Lagrange parameter that is dual to the constraint $\expec[\pmQ_T]{A_T} = a$. By solving the maximisation above, we obtain a driven process parameterised by $k$, which is equivalent to the one obtained by solving the constrained minimisation in \eqref{eqvar1} for a given $a$ such that $\lambda'(k)=a$ or $I'(a)=k$, consistent with the LF transform in \eqref{eqlf1} connecting $I(a)$ and $\lambda(k)$ \cite{chetrite2014}.

In practice, both optimisation problems are solved by considering \textit{ergodic} controlled processes, which means in the context of diffusions that the drift $F$ is changed to a new drift $\pF$ that makes the corresponding process $\pX_t$ ergodic. With this new drift, the evolution of $\pX_t$ is thus described by 
\begin{equation}
    \label{eq:controlSDE}
    d\pX_t = \pF(\pX_t)\,dt + \sigma\,dW_t.
\end{equation}
Alternatively, we can express this process as a perturbation of the original process by writing the controlled drift as $\pF = F + \Delta \pF$. The noise in all cases has to stay the same in order for the KL divergence to be defined.

The use of ergodic controls implies that $\expec[\pmQ_T]{A_T}$ converges as $T \to \infty$ to a constant $\pa^*$, given as in \eqref{eqstata1} in terms of the stationary density $\pp^*$ and stationary current $\pJ^*$ associated with $\pF$. Similarly, the term $D(\pmQ_T \Vert \pmP_T)/T$ converges to a stationary expectation relative to $\pX_t$ given by
\begin{equation}
    d(\pF \Vert F)  = \frac{1}{2}\int_{\reals^n} \pp^*(x)\,\Delta \pF(x)\cdot D^{-1} \Delta \pF(x)\,dx.
\end{equation}
Consequently, instead of obtaining $I(a)$ and $\lambda(k)$ by solving variational problems in path space which involve the long-time limit, we can optimise directly in the space of control drifts without that limit using
\begin{equation}
    \label{eqvarLD1}
    I(a) = \min_{\pF:\pa^*=a} d(\pF \Vert F)
\end{equation}
and
\begin{equation}
    \label{eqvarLD2}
    \lambda(k) = \max_{\pF} \left\{k \pa^* -d(\pF \Vert F)\right\}.
\end{equation}

Similar variational formulas are found for discrete-time Markov chains and continuous-time jump processes. What changes for each is the expression of the KL divergence rate and how the stationary value of $A_T$ is defined and calculated, as we explain in Sec.~\ref{secapps}. For instance, in the case of a Markov chain evolving in discrete time, this rate is expressed as a function $d(\pPi \Vert \Pi)$ involving the transition matrix $\Pi$ of the original Markov chain and a new transition matrix $\pPi$ describing a controlled Markov chain that is guided to a new stationary value of $A_T$.
The driven process obtained by maximising or minimising this rate, as above, has the same interpretation as in the diffusion case: it corresponds to the optimally-controlled ergodic Markov process that achieves a given fluctuation of $A_T$ as its typical value, ``optimality'' being defined in terms of the KL divergence, which plays the role of a control cost or learning loss.

The driven process is also known to correspond more physically to the Markov process that reproduces in the long-time limit the ensemble or subset of trajectories of $X_t$ associated with the event $A_T = a$ \cite{chetrite2014}. In this sense, it provides insights as to how fluctuations arise from specific trajectories, which are described by a modified process involving a modified drift or modified transition probabilities.

Naturally, since this process realises a value $A_T=a$ in a typical way, it is a good candidate for performing an importance sampling of $P(A_T=a)$, from which $I(a)$ can be estimated from the limit in \eqref{eqratelim1} \cite{touchette2011}. The advantage of the variational approach based on \eqref{eqvarLD1} is that $I(a)$ is obtained directly without having to first estimate $P(A_T=a)$. This distribution is known to have the large deviation form in \eqref{eqldq1}, and this fact is used directly to express $I(a)$ in terms of a KL divergence. In this sense, the driven process is used as an importance sampler to steer the process into rare regions where a control cost rather than a probability is estimated. This makes numerical methods based on the variational approach more efficient than standard importance sampling.

\section{The \ldstop method}
\label{secstochopt}

The \ldstop method is designed to solve the two variational problems in \eqref{eqvarLD1} and \eqref{eqvarLD2} yielding $I(a)$ and $\lambda(k)$, respectively. The idea of the method is to represent the controlled process $\pX_t$ using a set of parameters, denoted collectively by $\theta$, and to gradually change these parameters using a steepest descent (or ascent) method based on simulations of $\pX_t$ to reach the optimal process corresponding to the driven process. We describe in this section all the ingredients involved in the method. In practice, we consider the unconstrained maximisation in \eqref{eqvarLD2} to obtain $\lambda(k)$, since it is easier to solve than the constrained minimisation in \eqref{eqvarLD1}. Moreover, for explaining the method, we consider $X_t$ to be a diffusion process, unless stated otherwise. The modifications required for treating discrete-time Markov chains and jump processes are explained in Sec.~\ref{secapps}.

\subsection{Algorithm and estimators}

The pseudocode of the \ldstop method is shown in Algorithm~\ref{alg:ldstop}. The main steps of the method are as follows:
\begin{enumerate}
    \item Choose a representation for the control perturbation $\Delta \pF_\theta$ with parameters $\theta$ that is sufficiently flexible to represent the original and driven processes.
    
    \item Simulate the control process $\pX_t$ with $\pF_\theta=F+\Delta \pF_\theta$ over a time horizon $[0,T]$ and compute the observable $A_T$ in \eqref{eqobs1}, which serves as an estimator of the typical value $\pa^*$. To make sure that the estimation is accurate, the final time $T$ should be larger than the relaxation time of $\pX_t$, and the initial state $\pX_0$ should be in the region of the state space where the stationary distribution is mostly concentrated.
    
    \item From the same trajectory, compute the following estimator of the KL divergence rate:
    \begin{equation}
    K_T = \frac{1}{2T}\int_0^T \Delta \pF_\theta(\pX_{t}) \cdot D^{-1} \Delta \pF_\theta(\pX_{t}) dt.
    \end{equation}
    This estimate is computed numerically with a chosen discretisation step $\Delta t$. As for $A_T$, the conditions on $T$ and $\pX_0$ above ensure that $K_T$ is close to its ergodic limit.
    
    \item Using these estimates, define the loss $L_T$ derived from the unconstrained maximisation in \eqref{eqvarLD2} for a given value $k$, that is,
    \begin{equation}
    \label{eqloss}
    L_T = K_T - k A_T.
    \end{equation}
    Note that we change the sign compared to \eqref{eqvarLD2} because we want to minimise the loss, following the conventional usage in machine learning.
    
    \item Compute the gradient $\nabla_\theta L_T$ and use it to update the parameters in the direction that decreases the loss at the fastest rate. Note that this is an estimate of the gradient, not the true gradient of the expected loss associated with the variational problem in \eqref{eqvar2}.
    
    \item Repeat Steps 2 to 5 for $N$ epochs or until the parameters converge to the optimal ones, denoted by $\theta_k^*$.
    
    \item Use the last values of $A_T$, $K_T$ and $L_T$ as estimators of $I(a)$ and $\lambda(k)$. The estimator of the rate function at the value $A_T$ is $K_T$, whereas the estimator of the SCGF at the value $k$ is $-L_T$.
    
    \item Repeat the optimisation for various values of $k$, each yielding estimates for different points of the rate function and SCGF.
    The order of these optimisations is chosen to accelerate convergence through transfer learning.
\end{enumerate}

\begin{figure}[t]
\centering
\begin{minipage}{0.7\linewidth}
\begin{algorithm}[H]
    \caption{\ldstop}
    \label{alg:ldstop}
    \KwInput{$F$, $\sigma$, $\Delta \pF_\theta$, $\theta$, $\pX_0$}
    \KwParam{list or grid $kvals$ of $k$ values, $\Delta t$, $T$, $N$, $\alpha$}
    \For{$k \in kvals$}{
        \For{$N$ epochs}{
            generate trajectory $(\pX)_{t\in[0,T]}$ from $\pX_0$ using $\pF_\theta=F+\Delta \pF_\theta$\;
            compute $A_T$ and $K_T$ for that trajectory\;
            $L_T \leftarrow K_T - k A_T$\;
            compute $\nabla_\theta L_T$\;
            $\theta \leftarrow \theta - \alpha \nabla_\theta L_T$\;
            $\pX_0 \leftarrow \pX_T$\;
        }
        $a \leftarrow $ last $A_T$\;
        $I(a) \leftarrow $ last $K_T$\;
        $\lambda(k) \leftarrow $ last $-L_T$\;
        $\theta_k^* \leftarrow \theta$\;
    }
    
    \KwRet all estimates of $I(a)$ and $\lambda(k)$\;
\end{algorithm}
\end{minipage}
\end{figure}

We describe each of these steps in more detail in the next subsections. The crucial steps are the calculation of the gradient and the repetition of the simulations for different $k$ values, which can be done sequentially over a grid of points using transfer learning.

\subsection{Representation of the controlled dynamics}

The choice of representation for the controlled dynamics is vital to the performance of the optimisation method as a sufficiently expressive representation allows the controlled dynamics to converge to those corresponding to the driven process. In the case that a suboptimal representation is used, the control process will instead converge to the process closest to the driven process within the chosen representation class.\footnote{In either case, for simplicity, we assume that the chosen representation induces a loss landscape with properties allowing for convergence to the global minimum.} The final estimates for the rate function and SCGF, respectively, will in this case give upper and lower bounds to the true functions, since they are derived from the minimisation in \eqref{eqvarLD1} and maximisation in \eqref{eqvarLD2}.

Naturally, one may be unaware of the inflexibility of any chosen representation. A simple test is to use the converged dynamics as an importance sampler to estimate the rate function and SCGF.
If these estimates give tighter upper and lower bounds to the large deviation functions, the representation must be limited in its expressive power.

A wide range of function approximators is available to represent the control drift, including polynomials, rational functions, Fourier series, and neural networks, the latter being particularly attractive due to their strong expressive capabilities. Note that, in practice, it is more convenient to parameterise the perturbation $\Delta\pF$ than the drift $\pF$ itself. This is because the control process is then assured to represent the original process at $k=0$ by selecting the parameters such that $\Delta\pF=0$.

\subsection{Sequential \texorpdfstring{$k$}{k} values}

Each optimisation for a given value $k$ yields a point estimate of the rate function and SCGF. In our experience, to cover many values of $k$, it is best to start with $k = 0$, and to then move incrementally from this value in steps of $\Delta k$. For $k=0$, we know that both the rate function and SCGF are equal to zero, so starting the optimisation from this value ensures the correctness of the implementation and the adequacy of the selected parameters yielding $\Delta\pF_{\theta^*_0}=0$.

Once this first value is covered, we sequentially increment $k$ by a small step $\Delta k$, optimising for each value until a predetermined maximum value $k_{\max}$ is reached. This allows for transfer learning in the dynamics between each optimisation, wherein the parameters for a given $k$ value optimisation are initialised to those found for the previous optimisation. Since the $k$ values differ by a small increment, the parameters are initialised close to optimal ones for the current optimisation, and fewer epochs are required to converge than if the parameters were initialised differently.
Using a smaller increment $\Delta k$, as is likely to be done for applications wherein the large deviation functions are unknown, increases the advantageous effect of transfer learning.

This covers positive values of $k$. To cover negative values, we repeat the process by returning to $k=0$ and perform the optimisation in the same way with transfer learning by sequentially decreasing $k$ by $\Delta k$ until a set minimum value $k_{\min}$ is reached. Since the positive and negative loops are independent, one can easily run them in parallel.

In many applications, the domain of the SCGF is bounded, so we need to limit the bounds of $k$ properly. While these bounds may be unknown a priori, they are readily identifiable, as the variational representation of $\lambda(k)$ has no solution when $\lambda(k)=\infty$, leading in practice to a loss that continues decreasing in the optimisation.

An attractive alternative to the sequential scheme is to optimise the entire set of $k$ values concurrently, including $k$ in the input of the control drift \cite{yan2022}. In each epoch, multiple trajectories would be simulated, each for different $k$ values, and a single global loss would be constructed from their individual losses. However, the effect of transfer learning improves the performance of the method so substantially that we believe the sequential regime to be preferable.

Another possible modification is to specify a list of $k$ values that are not evenly spaced, or to use a varying increment size. This would be advantageous near dynamical phase transitions, where a small change in $k$ results in a much larger change in the related observable value and optimal control dynamics.

\subsection{Stochastic optimisation}

In each of the $N$ epochs, the parameters are updated through the stochastic gradient descent step according to
\begin{equation}
    \theta_{i+1} = \theta_{i} - \alpha_i \nabla_\theta L_T,
\end{equation}
where $\theta_i$ is the parameter set and $\alpha_i$ is the learning rate in the $i^{\text{th}}$ epoch.
This modifies each of the parameters in the direction that decreases the loss at the fastest rate.

The learning rate $\alpha_i$ can be determined through a variety of methods. Many established optimisation methods, such as stochastic gradient descent with momentum, Adagrad, Adam, and RMSprop, are implemented in most machine learning packages, such as PyTorch. These methods adjust the learning rate over the epochs through some specified mechanism, allowing the parameters to converge. For the applications in the next section, we have found that Adagrad \cite{duchi2011} performs well. Adagrad adaptively calculates an individual learning rate for each of the parameters in $\theta$ from the magnitudes of their components in $\nabla_\theta L_T$ over the previous epochs. One need only choose the initial learning rate $\alpha_0$. Other optimisers may perform better depending on the form of the parameterisation.

\subsection{Gradient computation}
\label{secgradient}

There are essentially three options available for computing the gradient of the loss, depending on the type of Markov processes considered. The main point to note for understanding each option, discussed next, is that the parameters $\theta$ enter not only in the expression of the loss (in the case of diffusions, via $\Delta\pF_\theta$), but also in the generation of the trajectory $(\pX_t)_{t\in[0,T]}$ used to estimate the loss with $L_T$. When computing the gradient, both of these dependencies must be taken into account. 

\begin{itemize}
\item \textbf{Stochastic automatic differentiation}: For diffusions, it is normal to use a representation $\Delta\pF_\theta$ (e.g., a neural network) that is differentiable in $\theta$, which means that the generation of a trajectory $(\pX_t)_{t\in [0,T]}$ with an SDE solver is also differentiable in $\theta$. In this case, the gradient estimator $\nabla_\theta L_T$ can be computed directly using automatic differentiation, via the \texttt{grad} operator available in PyTorch, TensorFlow or JAX, which applies the chain rule to the full sequence of operations involved in the computation of $L_T$, all the way down to the generation of the underlying trajectory. This can be done efficiently in practice if the diffusion considered is low-dimensional and the simulation time $T$ is not too large. When the \texttt{grad} operator is applied for the first time, a computational graph describing the chain rule applied to $L_T$ is constructed internally, and used afterwards in future calls of this operator. From that point, the complexity of computing the gradient is the same as that of computing $L_T$ itself.
    
Automatic differentiation can also be used for other types of processes if the loss and the generation of the trajectories underlying the loss are differentiable in $\theta$. As an example, we consider in the next section  sums of independent random variables that can be differentiated sample-wise with respect to some parameters. Many machine learning packages implement this differentiable generation of random variates, based on differentiable transformations of simpler random variables (e.g., uniform or standard normal). The \texttt{rsample} method from the PyTorch \cite{paszke2019} library, for example, makes this approach particularly simple and convenient.

\item \textbf{Stochastic adjoint sensitivity method}: The generation of the computational graph underlying the automatic gradient becomes computationally prohibitive when considering complex (e.g., many-particle) diffusion processes. For those, it was suggested in \cite{yan2022} to use the adjoint state method instead to compute $\nabla_\theta L_T$. This method, developed in \cite{li2020}, is a stochastic generalisation of the adjoint sensitivity method for computing gradients of deterministic systems, which uses the ``forward'' trajectory $(\pX_t)_{t\in[0,T]}$ and the ``backward'' version of this trajectory to estimate the gradient (see Appendix C of \cite{yan2022} for further details). The TorchSDE library, which accompanies \cite{li2020}, implements various differentiable SDE solvers and has the option to compute gradients using the adjoint method. Note that this method is only available for SDEs.

\item \textbf{Score gradient estimator}: Processes with discrete states cannot in general be dealt with using automatic differentiation because they are not generated from methods or algorithms that are differentiable with respect to their parameters. An alternative, for those, is to compute the gradient using the \textit{score method}, which replaces the gradient of an expectation by a new expectation featuring an additional term, called the score \cite{williams1992,sutton2018,arampatzis2016,engel2023}. 
    
To understand the basis of this method in a simple way, consider computing the gradient
\begin{equation}
    \nabla_\theta  \expec[p_\theta]{f(X)},
\end{equation}
where $X$ is a real random variable distributed according to some parameterised density $p_\theta$. Writing the expectation explicitly and introducing the gradient inside the expectation gives
\begin{equation}
    \nabla_\theta \int_\reals p_\theta(x) f(x)\, dx = \int_\reals p_\theta(x) f(x)\, \nabla_\theta\ln p_\theta(x) dx.
\end{equation}
Therefore,
\begin{equation}
    \nabla_\theta \expec[p_\theta]{f(X)} =  \expec[p_\theta]{f(X) \nabla_\theta\ln p_\theta(X)}.
\end{equation}
    
From this result, we see that the gradient can be estimated by
\begin{equation}
    \nabla_\theta  L_n=\frac{1}{n}\sum_{i=1}^n f(X^{(i)}) \nabla_\theta\ln p(X^{(i)}),
\end{equation}
from a sample $(X^{(i)})_{i=1}^n$ of $n$ realisations of $X$. The difference with the automatic gradient method is that no derivatives of the sample values are taken; the gradient is only applied to the log-density, which is assumed to be differentiable in $\theta$. This gradient is the score function.
    
We explain in Sec.~\ref{secappIID1} how this estimator transposes to the loss underlying the large deviations of sample means of independent variables, as well as Markov chains evolving in discrete time. For the latter, we find 
\begin{equation}
    \label{eqscoreMCnobase}
    \nabla_\theta L_T = L_T \sum_{t = 1}^{T} \nabla_\theta \ln \pPi_\theta(\pX_{t-1}, \pX_t),
\end{equation}
where $(\pX_t)_{t\in [0:T]}$ is a trajectory of the controlled Markov chain, generated with a parameterised transition matrix $\pPi_\theta$, and $L_T$ is the estimated loss for that process.
    
The score gradient is known in reinforcement learning as the REINFORCE gradient \cite{williams1992,sutton2018}. In practice, it is unbiased, but has a high variance, which can be reduced by subtracting a constant from $L_T$, called the baseline, which needs to be independent of the state of the process generated. For our purposes, we use a baseline computed from previous epochs, so as to consider the following gradient estimator:
\begin{equation}
     \label{eqscoreMCaveloss}
     \nabla_\theta L_T = (L_T-\bar{L})\sum_{t = 1}^{T} \nabla_\theta \ln \pPi_\theta(\pX_{t-1}, \pX_t),
\end{equation}
where $\bar{L}$ is the average loss over the previous epochs of the current $k$ value optimisation. Another form of baseline based on a partial loss is used in Sec.~\ref{secJP1} to further stabilise the variance of this estimator.\footnote{Many packages implement a function (e.g., \texttt{log\_prob} in PyTorch) returning the log of a probability distribution or density, which can also be used to get the log of a transition matrix.} 

\end{itemize}

\subsection{Averaging over many trajectories}

The accuracy of the estimators $A_T$, $K_T$ and $L_T$ is greatly improved by averaging them over $n$ independent trajectories, $(\pX_t^{(i)})_{t\in[0,T]}$, $i\in [1:n]$, as suggested in \cite{yan2022}. This is advantageous over increasing $T$, as the independent trajectories can be computed in parallel, whereas a single longer trajectory must be computed in series. Appendix~B of \cite{yan2022} contains a justification of this choice as well as calculations showing that the loss estimate's variance scales as $(nT)^{-1}$, so there is a direct tradeoff between increasing the number of trajectories and decreasing their duration.

The initial state $\pX_0$ of the parallel trajectories is initialised only for the first epoch of the $k = 0$ optimisation. Thereafter, we use the final states $\pX_T^{(i)}$ of the previous epoch as the initial state of the current epoch. This initialises the state closer to the stationary distribution of the control process, allowing faster convergence of the estimates and loss. Additionally, this transfer learning in state is applied between the optimisations of sequential $k$ values, excepting the first negative $k$ optimisation, in which we reinitialise $X_0$ according to the specified value used previously.

To compute the final estimates of the observable, rate function, and SCGF for the given $k$ value, we also average $A_T$, $K_T$ and $L_T$ over a number $M$ of previous epochs, choosing $N\gg M$ to ensure that the dynamics and estimates have converged. Another option is to use the converged dynamics in a new simulation from which to compute the estimates. This last simulation can be done at a lower computational cost, since the gradient computation is not needed.

\section{Applications}
\label{secapps}

We demonstrate in this section the \ldstop method. Various examples, chosen specifically for their simplicity, are used to illustrate important details of the method's implementation and use \cite{cloete2025}. The examples allow for a comparison of the \ldstop estimates to the true SCGF and rate function, since these are either known or can be approximated numerically with high precision. We also explain the modifications to the method required when considering processes other than diffusions. Most of the examples that we discuss are contained in a notebook available on GitHub; see the link in Footnote~\ref{github}.

\subsection{Sample means of independent random variables}
\label{secappIID1}

For our first application, we consider a sample mean of the form
\begin{equation}
    A_T = \frac{1}{T} \sum_{t=1}^{T} X_t   
    \label{eqsamplemean1}
\end{equation}
involving an integer number $T$ of independent and identically distributed (IID) random variables $(X_t)_{t\in [1:T]}$, distributed according to a density or mass function $p(x)$. The determination of the SCGF and rate function for this observable is a well-known problem in large deviation theory; see \cite{touchette2009,burenev2025} for reviews. The KL divergence rate entering in the variational formulas in \eqref{eqvarLD1} and \eqref{eqvarLD2} is given in this case by the relative entropy distance
\begin{equation}
    d(\pp \Vert p) = \sum_{x} \pp(x) \ln \frac{\pp(x)}{p(x)}
    \label{eqKLiid1}
\end{equation}
between some modified density $\pp$ and the original density $p$. (For continuous random variables, the sum is replaced by an integral.) The maximisation in \eqref{eqvarLD2}, determining the SCGF, can be carried out analytically, yielding the tilted density
\begin{equation}
    \pp_k(x) = \frac{e^{kx} p(x)}{\expec[p]{e^{kX}}}.
\end{equation}
This acts analogously to the driven process, in that the optimal process underlying a fluctuation $A_T=a$ of the sample mean is another IID sequence of random variables distributed according to $\pp_k$ with $k$ chosen according to $\lambda'(k)=a$.

The specific form of the exponential tilting is known in many cases, and we exploit this information in the following two examples. To find this tilting among a class of parameterised densities $\pp_\theta$, we use the \ldstop method, based on the following natural estimator of the KL divergence rate:
\begin{equation}
    \label{eq:IIDKL}
    K_T = \frac{1}{T}\sum_{t= 1}^{T} \ln \frac{\pp_\theta(\pX_t)}{p(\pX_t)},
\end{equation}
where $(\pX_t)_{t\in[1:T]}$ is a sample obtained from $\pp_\theta$. Depending on the nature of the random variables summed, we can minimise the loss $L_T=K_T-kA_T$ to find the SCGF using either the stochastic gradient or the score gradient. This is illustrated next.

\subsubsection{Exponential random variables}

We first consider the $X_t$'s in the sample mean to be exponentially distributed with expected value $\mu = 1$. Then the exponential tilting is known to be another exponential distribution with a different expected value, so we choose the controlled density $\pp_\theta$ to be another exponential distribution with expected value $\theta$. Other distributions can also be used, provided that they can approach the exponential one. For the latter, the generation of variates is continuous in $\theta$, so we also make use of automatic differentiation to compute the gradient of $L_T$.

\begin{figure*}[t]
\centering
\includegraphics[width=\textwidth]{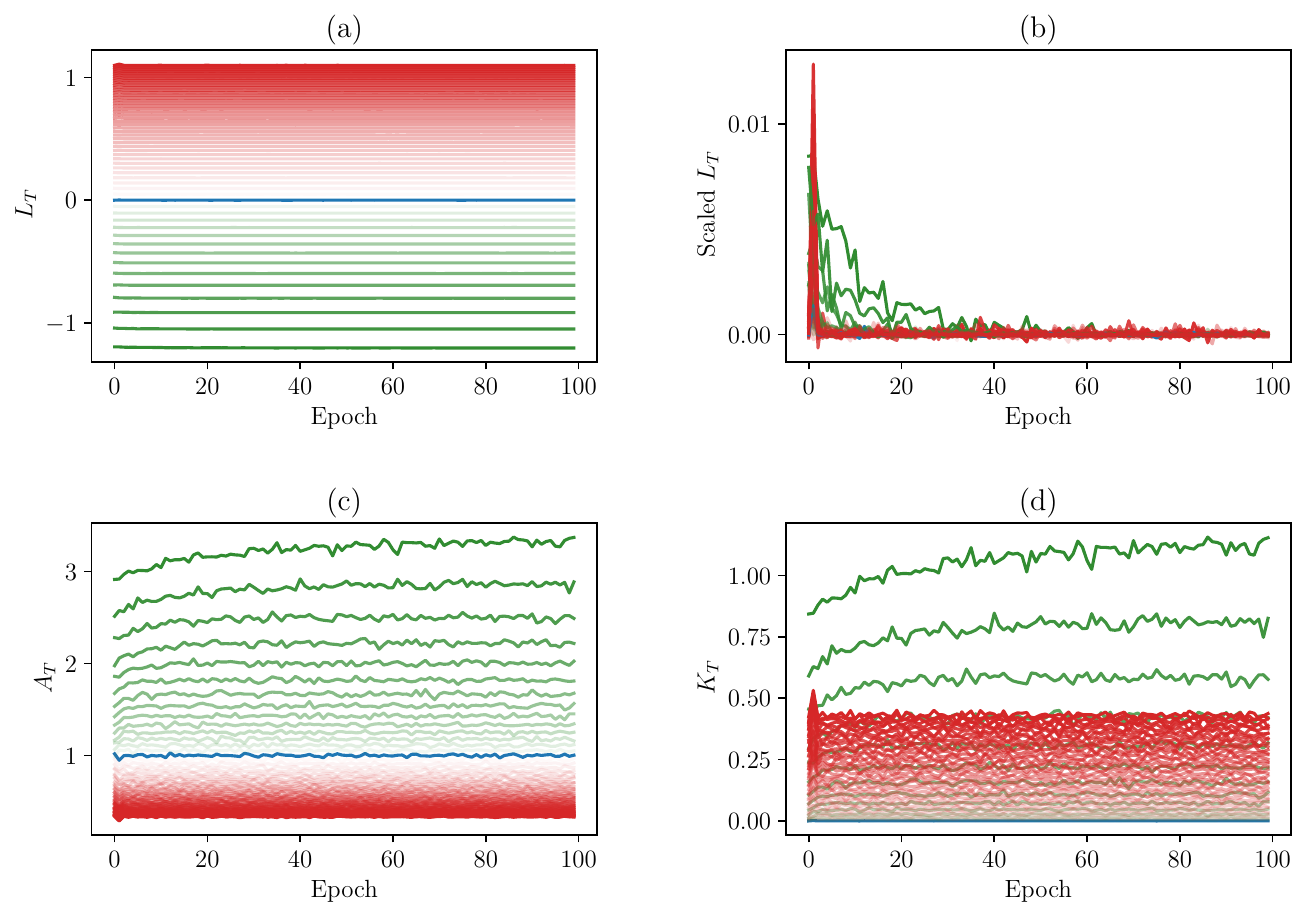}
\caption{Convergence of \ldstop for the exponential IID sample mean with expected value $\mu = 1$ using an exponential distribution for the control. In all the plots, $k = 0$ is shown as blue, the positive $k$ values are shown as green, whereas the negative $k$ values are shown as red. Moreover, the opacities of the non-zero $k$ value plots are scaled so that those corresponding to the largest magnitude $k$ values are completely solid, and those for smaller magnitude $k$ values are more transparent.}
\label{fig:IIDexpConvergence}
\end{figure*}

\begin{figure*}[t]
    \centering
    \includegraphics[width=0.875\hsize]{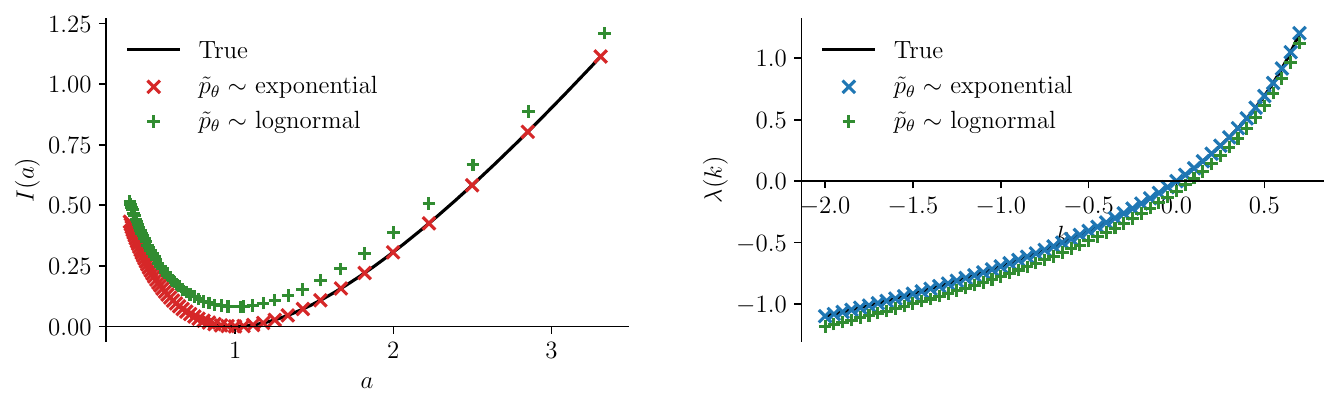}
    \caption{Rate function (left) and SCGF (right) for the exponential IID sample mean with expected value $\mu = 1$.
    Red and blue crosses denote the corresponding \ldstop estimates using an exponential distribution for the control, and green crosses denote those obtained using a lognormal distribution for the control.}
    \label{fig:IIDexpResults}
\end{figure*}

We show the convergence of \ldstop based on this gradient in Fig.~\ref{fig:IIDexpConvergence}. For the simulations, we used a grid of $k$ values equally spaced between $k_{\min} = -2$ and $k_{\max} = 0.7$ with $\Delta k = 0.05$. We also used $N = 100$ optimisation epochs, and a total of $T = 10\,000$ independent variates sampled, averaging the final $M = 25$ epochs to estimate the SCGF and rate function reported in Fig.~\ref{fig:IIDexpResults}. Adagrad with the initial learning rate $\alpha_0 = 0.05$ was employed to calculate the learning rates for each epoch. Finally, at $k=0$, we used $\theta = 1$ to match the original density.

The plot in Fig.~\ref{fig:IIDexpConvergence}(a) shows the loss $L_T$ over the epochs, the different colours and opacity representing different $k$ values (see caption). When plotting the loss in this way, it is generally not possible to observe the convergence of $L_T$. For this reason, we scale the losses in such a way that the average of their final $M$ values is equal to zero, as shown in Fig.~\ref{fig:IIDexpConvergence}(b). From the scaled losses, we can observe that the method converges after about $50$ epochs.

This convergence is also seen to some extent at the level of $A_T$ and $K_T$, shown in Fig.~\ref{fig:IIDexpConvergence}(c) and (d), respectively. The effect of transfer learning is visible in these plots, in the way that the estimates of the final positive $k$ value optimisation (the highest green plots) start at the same value that the previous $k$ value's estimates end (the second-highest green plots).

The final \ldstop estimates for the rate function and SCGF are shown in Fig.~\ref{fig:IIDexpResults}, where they are compared to their exact values obtained analytically. It can be seen that the estimates approximate the large deviation functions well. Additionally, to illustrate the effect of using a limited representation, i.e., a representation that does not include the optimal $\pp$, we show the final estimates obtained when using a lognormal control instead of an exponential one (using the same parameters as before). These clearly show that the \ldstop estimates yield an upper and lower bound to the rate function and SCGF, respectively, as discussed in Sec.~\ref{secstochopt}. These are the best possible estimates that can be obtained within the class of lognormal densities.

\subsubsection{Bernoulli random variables}

To illustrate the case of non-differentiable variates, we now consider the $X_t$'s to be Bernoulli-distributed with expected value $p = \frac{1}{2}$. The form of the exponential tilting in this case is known to be a Bernoulli distribution with a different expected value, so we represent $\pp_\theta$ now as a Bernoulli distribution with expected value $\theta$. The generation of random values from this distribution is not differentiable (it involves an `if' statement), so we cannot use automatic differentiation. As a result, we instead use the score gradient estimator with an average loss baseline, given by
\begin{equation}
    \label{eqscoreIID}
    \nabla_\theta L_T = \frac{1}{T} \sum_{t = 1}^{T} (L_t - \bar{L}) \nabla_\theta \ln \pp_\theta(\pX_t),
\end{equation}
where 
\begin{equation}
    L_t = \ln \frac{\pp_\theta(\pX_t)}{p(\pX_t)} - k \pX_t
\end{equation}
is the loss contribution of each $\pX_t$, and $\bar{L}$ is the average loss over the previous epochs in the current $k$ value optimisation. (See Appendix~\ref{secappend} for the derivation of this estimator.)

\begin{figure*}[t]
    \centering
    \includegraphics[width=0.875\hsize]{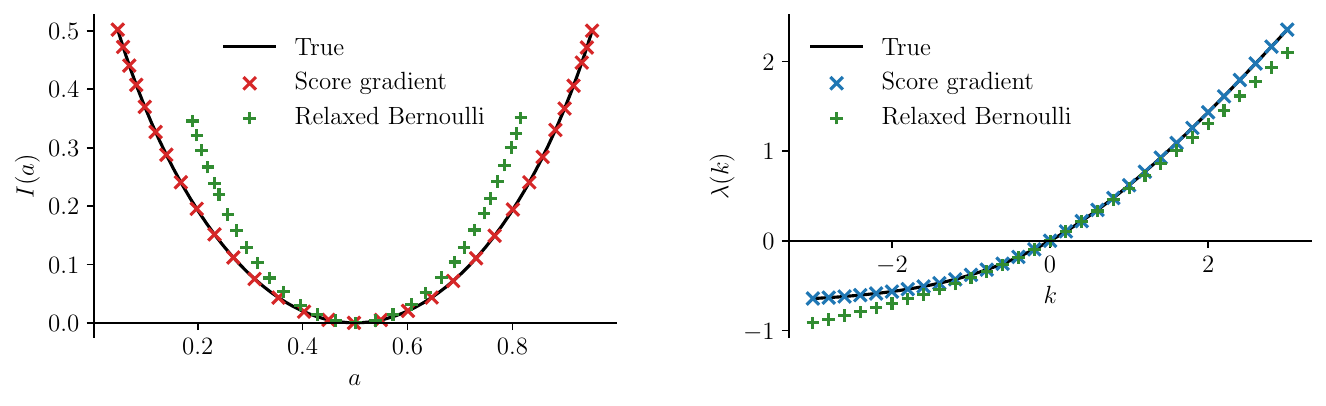}
    \caption{Rate function (left) and SCGF (right) for the Bernoulli IID sample mean with expected value $p = \frac{1}{2}$. Red and blue crosses denote the corresponding \ldstop estimates using the score gradient, and green crosses denote those obtained using stochastic automatic differentiation via the relaxed Bernoulli representation.}
    \label{fig:IIDbernResults}
\end{figure*}

The results of the \ldstop method obtained with this gradient estimator are shown in Fig.~\ref{fig:IIDbernResults}. For the simulations, we initialised the control parameter to be $\theta_0 = \frac{1}{2}$, and used $k_{\min} = -3$ and $k_{\max} = 3$ with $\Delta k = 0.2$ to generate the grid of $k$ values. A total of $T = 10\,000$ variables were generated using $\pp_\theta$ in each of $N = 25$ epochs per $k$ value. To compute the final estimates, the final $M = 10$ epochs were averaged. Finally, we used the Adagrad optimiser, as before, with the initial learning rate $\alpha_0 = 0.02$. As can be seen, the numerical results are in very good agreement with the exact results obtained analytically \cite{touchette2009}.

As an alternative to using the score gradient, one can use a continuous approximation of the Bernoulli distribution such as the relaxed Bernoulli to allow the use of stochastic automatic differentiation.
This has another parameter in addition to the expected value, called the temperature, which gives a measure of how closely it approximates the true discrete Bernoulli distribution.
We include in Fig.~\ref{fig:IIDbernResults} the results of \ldstop (using the same parameters as before) when using this relaxed Bernoulli with temperature equal to $0.1$, taking care to also replace the base distribution $p$ with a relaxed Bernoulli to make sure that the KL divergence is defined.

When using the relaxed Bernoulli distribution, we can see that there are significant errors, which come from two factors: (i) the fact that the relaxed Bernoulli does not exactly represent the tilted Bernoulli, and (ii) our replacement of $p$ with a relaxed Bernoulli, which changes the underlying sample mean. Both of these demonstrate the limitation of replacing discrete random variables with continuous ones.

\subsection{Discrete-time Markov chains}
\label{secMCld1}

We now move to applications involving discrete-time Markov chains $(X_t)_{t\in [1:T]}$, described by a time-independent transition matrix
\begin{equation}
    \Pi(x,y) = P(X_{t+1} = y \vert X_t = x),
\end{equation}
representing the probability of reaching a state $y$ from a state $x$ in one time step. For simplicity, we assume that the state $X_t\in\cX$ is discrete. For this type of processes, the observables of interest take the general form
\begin{equation}
    A_T = \frac{1}{T} \sum_{t = 1}^{T} g(X_{t-1}, X_t),
\end{equation}
where $g$ is some real function defined on $\cX\times\cX$. To find the large deviations of this observable, we use the KL divergence rate, which is known for ergodic Markov chains to take the form
\begin{equation}
    d(\pPi \Vert \Pi) = \sum_{x,y} \pP^*(x) \pPi(x,y) \ln\frac{\pPi(x, y)}{\Pi(x, y)},
\end{equation}
where $\pP^*(x)$ is the stationary distribution of the controlled Markov chain with transition matrix $\pPi$ \cite{chetrite2014}. This KL rate is estimated for a parameterisation $\pPi_\theta$ of this chain by computing
\begin{equation}
    K_T = \frac{1}{T}\sum_{t = 1}^{T} \ln\frac{\pPi_\theta(\pX_{t-1}, \pX_t)}{\Pi(\pX_{t-1}, \pX_t)},
    \label{eqKTMC1}
\end{equation}
from a trajectory $(\pX_t)_{t\in[1:T]}$ generated from $\pPi_\theta$. 

A simple way to parameterise $\pPi_\theta$ is to use another matrix of the same dimension as the original with the matrix elements being the parameters $\theta$. To ensure relative continuity between the processes, no transitions should be possible in the controlled process that are not possible in the original process. In practice, we can enforce this constraint by setting the corresponding elements of $\pPi_\theta$ to 0. Furthermore, the rows should be normalised to sum to 1, so that $\pPi_\theta$ defines a valid transition matrix, which can be achieved by dividing each row by its sum.\footnote{For Markov chains with an infinite state space, the original process will have a local way of specifying the transition matrix. The control transition matrix can then be represented similarly.}

Another natural representation option is to have the parameters be the non-zero elements of $\pPi_\theta$, with the final non-zero element of each row being computed from the others so that the row sums to one. This parameterisation is advantageous in that it does not require normalisation or enforcing the relative continuity constraint. However, most machine learning frameworks do not support constructing a transition matrix whose entries remain automatically linked to the underlying parameters. As a result, the transition matrix must be reconstructed from the updated parameters after each optimisation step, making the  former parameterisation preferable, especially in Markov chains with large state spaces.

Note that, in all cases, the generation of a trajectory from $\pPi_\theta$ is not a differentiable procedure with $\theta$, since it involves a random selection of discrete states, as in the case of the sample mean of Bernoulli random variables. For this reason, the minimisation of $L_T$ must be performed with the score gradient, with a baseline to mitigate the variance, as described in Sec.~\ref{secgradient} and Appendix~\ref{secappend}.

\subsubsection{Random walk on a random graph}

\begin{figure*}[t]
    \centering
    \includegraphics[width=0.45\hsize]{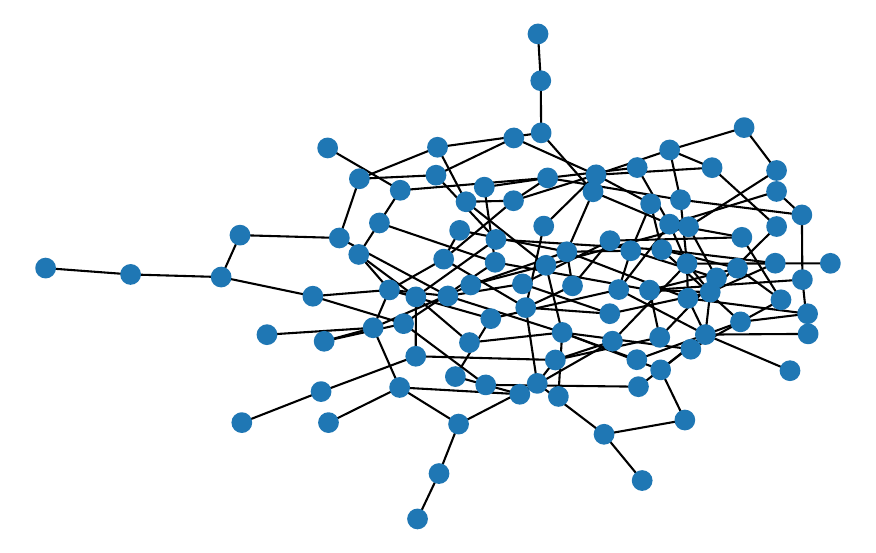}
    \caption{Erd\H{o}s--R\'enyi graph with $100$ vertices and $0.025$ probability of edge creation.}
    \label{fig:graph}
\end{figure*}

To test \ldstop for Markov chains, we consider the large deviations of the mean degree visited by an unbiased random walk on an Erd\H{o}s--R\'enyi graph \cite{bacco2016, coghi2019}. We choose this application, as it exhibits a dynamical phase transition, which is known to be troublesome for the cloning algorithm \cite{yan2022} and, to a lesser extent, also for spectrum-based approaches, such as the adaptive power method (APM) studied recently \cite{coghi2023}. 

This random walk is defined as follows. We consider the undirected random Erd\H{o}s--R\'enyi graph, shown in Fig.~\ref{fig:graph}, consisting of 100 vertices, which have been attached pairwise with probability $0.025$, and run a random walk on the vertices by selecting a connected node at random to jump to, so the transition probabilities are given by 
\begin{equation}
\Pi_{x,y} = \frac{A_{x,y}}{\deg(x)},
\end{equation}
where $A$ is the adjacency matrix, with element $A_{x,y} = 1$ if an edge exists between vertices $x$ and $y$ and $A_{x,y} = 0$ otherwise, and $\deg(x)$ is the degree of node $x$. This random walk is ergodic if the graph is connected. The observable that we investigate is the average vertex degree 
\begin{equation}
    A_T = \frac{1}{T} \sum_{t = 1}^{T} \deg(X_t),
\end{equation}
visited by the random walk over a time $T$.

To represent the controlled transition matrix $\pPi_\theta$, we use a 100 by 100 matrix initialised such that it is equal to $\Pi$. Each element, even those equal to $0$, is a parameter that could change.
To enforce the constraint that no transitions are possible in the controlled process that are not possible in the original process, we apply an  element-wise multiplication of $\pPi_\theta$ by the adjacency matrix $A$. We then clamp the elements to be non-negative, that is, we set the negative elements equal to 0, and normalise the rows to sum to one.\footnote{Many packages will automatically normalise the input when defining a multinomial distribution, making manual normalisation unnecessary.}

With this representation, we ran the \ldstop method from $k_{\min} = -0.5$ to $k_{\max} = 0.5$ in steps of $\Delta k = 0.025$, using the Adagrad optimiser learning rate initialised to $\alpha_0 = 0.025$ over $N = 100$ epochs. For each epoch, we computed estimates by simulating in parallel a total of $n = 1\,000$ independent trajectories up to a final time $T = 150$, and then averaged these estimates over the last $M = 20$ epochs to get the large deviation functions.

\begin{figure*}[t]
    \centering
    \includegraphics[width=0.875\hsize]{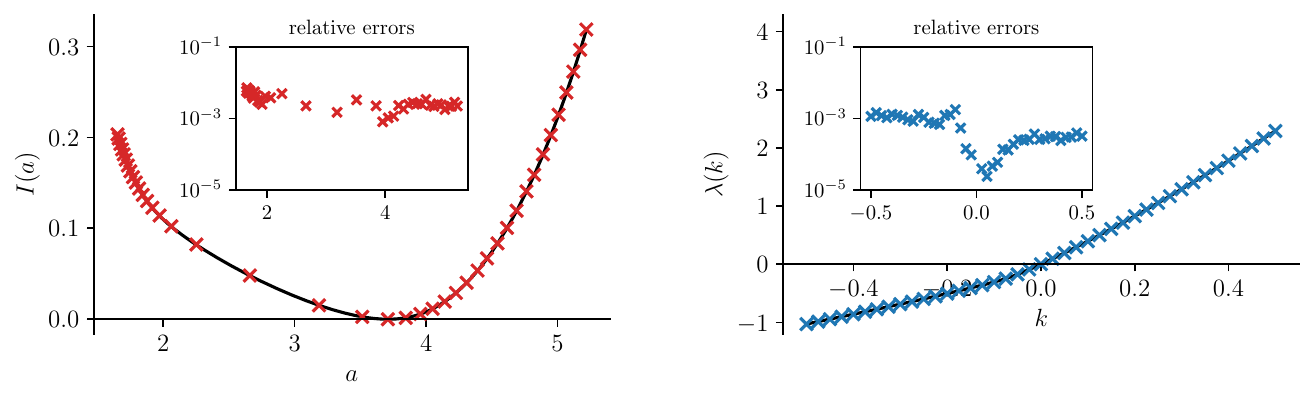}
    \caption{Rate function (left) and SCGF (right) for the mean vertex degree of an unbiased random walk on an Erd\H{o}s--R\'enyi graph. Crosses denote the corresponding \ldstop estimates, and the insets show the relative errors.}
    \label{fig:MCgraph}
\end{figure*}

The final results, shown in Fig.~\ref{fig:MCgraph}, accurately recover the true functions, computed numerically using an eigenvalue solver. The insets show that the relative errors are between $10^{-2}$ and $10^{-5}$, even in the region of the dynamical phase transition, which corresponds to the linear part of $I(a)$ or the kink in $\lambda(k)$. This is remarkable. Compared to the results obtained using the APM in \cite{coghi2023} (see their Fig.\ 4 for instance), our method does not display any sticking or instability in the region of the dynamical phase transition. This is likely due to the use of transfer learning, which ensures that the solution found for a given $k$ value is used as a seed for the next value $k+\Delta k$, and the relative stability of solving an optimisation problem with self-averaging loss and gradient estimators.

For the same system, we tried another score-based gradient estimator suggested by actor-critic methods in reinforcement learning \cite{sutton2018}. The difference with the previous REINFORCE estimator is that the parameters can be updated after each transition instead of only after simulating an entire trajectory \cite{cloete2025}. The results obtained are as accurate, but converge faster than with the REINFORCE estimator; see \cite{cloete2025} for the details.

\subsection{Markov jump processes}
\label{secJP1}

For the next application, we consider Markov jump processes, described through the generator matrix 
\begin{equation}
    G(x,y) = W(x,y) - \nu(x)\delta_{x,y},
\end{equation}
where $W(x,y)$ are the transition rates, 
\begin{equation}
\nu(x) = \sum_{y \neq x} W(x,y)
\end{equation}
are the escape rates, and $\delta_{x,y}$ is the Kronecker delta. We denote the times at which jumps occur as $t_1, t_2, t_3, \ldots$, and the time spent in state $X_{t_i}$, called the residence or sojourn time, by $\tau_i = t_{i+1} - t_i$. The residence times are known to be exponentially distributed because of the Markov property.

The general observable that we study for these processes includes, similarly to the case of diffusions, an integral part and a jump part, expressed as
\begin{equation}
    A_T = \frac{1}{T} \int_{0}^{T} f(X_t)\,dt + \frac{1}{T} \sum_{i = 1}^{m} g(X_{t_{i-1}}, X_{t_i}),
\end{equation}
where $m$ denotes the number of transitions that occur before the final time $T$. The KL divergence rate that determines the large deviations of this observable now takes the form
\begin{equation}
    d(\pW \Vert W) = \sum_{x,y} \pP^*(x) \pW(x,y) \ln\frac{\pW(x,y)}{W(x,y)} - \sum_x \pP^*(x)\left[\pnu(x) - \nu(x)\right],
    \label{eqJPKL1}
\end{equation}
where $\pP^*(x)$ is the stationary distribution of the controlled process with transition rates $\pW(x,y)$, to which are associated the escape rates $\pnu(x)$.

Trajectories of jump processes are simulated by generating a sequence of jumps according to the transition matrix $\pPi(x,y) = \pW(x,y)/\pnu(x)$, and by spacing these jumps in time according to a sequence of residence times $\tau_i$ generated using $\pnu(x)$. As a result, it is natural to represent the controlled process by parameterising separately the transition matrix as $\pPi_{\theta_1}$ and the escape rate vector as $\pnu_{\theta_2}$ with two parameter sets $\theta_1$ and $\theta_2$. The controlled generator can then be reconstructed through $\pW_\theta(x,y) = \pnu_{\theta_2}(x) \pPi_{\theta_1}(x,y)$, with $\theta=(\theta_1,\theta_2)$ as the full parameter set. For representing the transition matrix, we follow what we did for Markov chains. As for the escape rates, one can have the parameters be the escape rates themselves, with the only constraint being that these are strictly positive. This can be enforced by taking the elements to be the exponential of a real parameter, or by setting the non-positive elements equal to some small positive number.

Note that, because of the simulation method used, we must specify the number of jumps $m$ instead of the final time $T$ for each trajectory in each epoch. (One could instead simulate up to a final time by checking each trajectory's current time $t$.) Each trajectory thus has a unique final time $T^{(i)}$, which must be used to compute each trajectory's respective observable and the KL divergence rate estimate. Based on the form of the KL rate in \eqref{eqJPKL1}, the estimator that we use for a single trajectory is
\begin{equation}
    K_T = \frac{1}{T} \sum_{i = 1}^{m} \left( \ln\frac{\pW_\theta(\pX_{t_{i-1}}, \pX_{t_i})}{W(\pX_{t_{i-1}}, \pX_{t_i})} - \tau_i\left[\pnu_{\theta_2}(\pX_{t_{i-1}}) - \nu(\pX_{t_{i-1}})\right] \right).
\end{equation}

The gradient of $L_T$ based on this KL estimator is computed separately along the two sets of parameters. On the one hand, for computing gradients relative to $\theta_2$, we use automatic differentiation, since the escape times are generated in a differentiable way, as in our first IID example, from an exponential density. On the other hand, for computing gradients in $\theta_1$, we make use of the score gradient, since this part relates to generating a Markov chain in discrete time. Specifically, we use the score gradient estimator in \eqref{eqscore}, which includes a baseline and a sum of partial losses truncated to the time $\lfloor T/2\rfloor$. For comparison, we also computed the same gradient, but without truncating the sum, so the sum is performed up to $T$ rather than $\lfloor T/2\rfloor$. 

To test \ldstop method with these gradients, we consider a minimal jump process with a state space $\cX = \{0,1,2\}$ and transition rates $W(x,y) = \frac{1}{2}$ for all distinct states $x$ and $y$. The escape rates are thus $\nu(x) = 1$ for all states. The observable that we consider is the mean state 
\begin{equation}
    A_T = \frac{1}{T}\int_{0}^{T} X_t\,dt.
\end{equation}

\begin{figure*}[t]
    \centering
    \includegraphics[width=0.875\hsize]{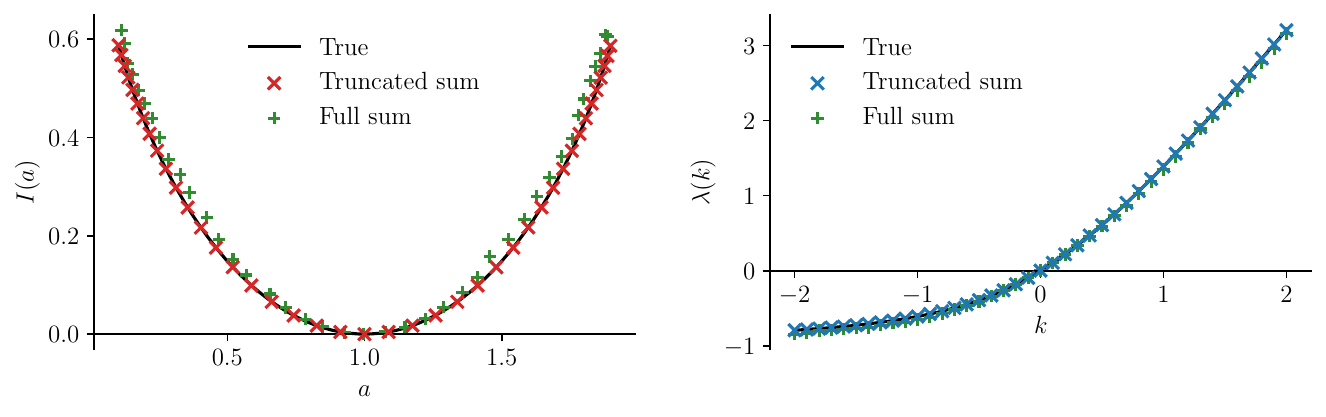}
    \caption{Rate function (left) and SCGF (right) for the three-state jump process sample mean. Red and blue crosses denote the corresponding \ldstop estimates using the truncated sum score gradient, and green crosses denote those obtained using the full sum score gradient.}
    \label{fig:MJPthreeResults}
\end{figure*}

The SCGF and rate function obtained for this process and observable are shown in Fig.~\ref{fig:MJPthreeResults}. The results obtained with the truncated score gradient estimator are shown in red and blue, whereas those obtained with the untruncated (full) score gradient estimator are shown in green. Both are compared with the exact rate function obtained by diagonalisation \cite{burenev2025}. To estimate the large deviation functions, we used a grid of $k$ values between $k_{\min} = -2$ and $k_{\max} = 2$ separated by $\Delta k = 0.1$. Moreover, we used $N = 100$ epochs, simulating $n = 1\,000$ parallel trajectories containing $m = 50$ jumps in each. The Adagrad learning rate was initialised to $\alpha_0 = 0.05$, and the final $M = 50$ epochs' estimates were averaged to obtain the final large deviation functions.

\begin{figure*}[t]
    \centering
    \includegraphics[width=0.875\hsize]{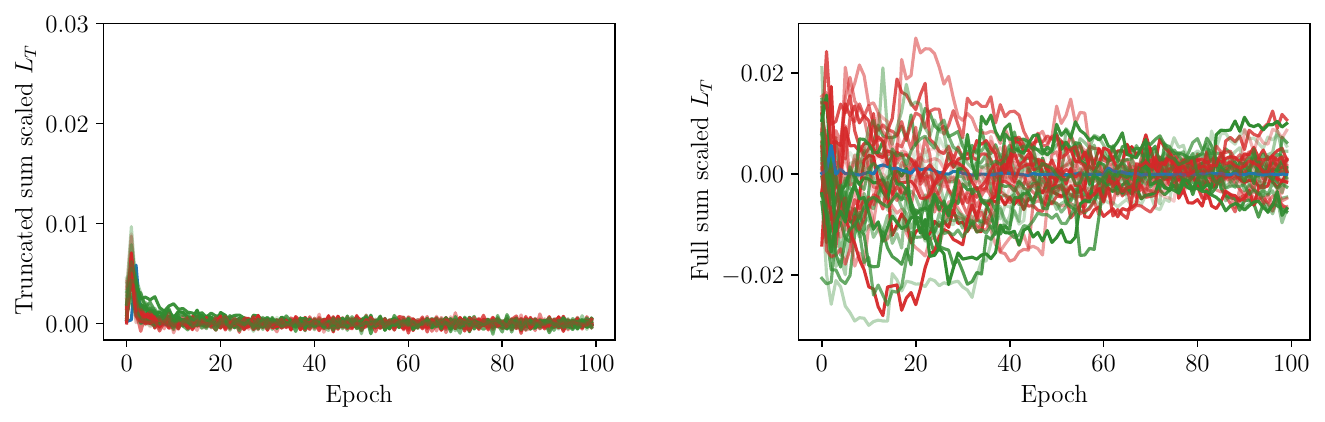}
    \caption{Scaled loss of \ldstop using the truncated sum score gradient (left) and using the full sum score gradient (right) for the three-state jump process sample mean.}
    \label{fig:MJPthreeLoss}
\end{figure*}

From the results, shown in Fig.~\ref{fig:MJPthreeResults}, we see that the truncated score gradient estimator performs better than the full score gradient, particularly for values of $a$ further from the typical value $a^* = 1$. The difference in performance between these two estimators is more clearly seen in Fig.~\ref{fig:MJPthreeLoss}, showing that the (scaled) loss converges within $30$ epochs for the truncated score gradient estimator (left panel). By contrast, the scaled loss does not seem to converge in the full score gradient estimator, shown in the right panel, and for some optimisation steps even increase. The higher variance in the second case is caused by partial loss terms coming near the final times close to $t=T$, which are noisier because of their shorter integration times. This observation is actually what led us to the idea of truncating in time the score gradient estimate. To confirm our results, we also used the score gradient estimator in \eqref{eqscoreMCaveloss}, which is not truncated in time, and found results that are similar to the ones found for the truncated partial score gradient in \eqref{eqscore}. In practice, we recommend using one or the other.

\subsection{Markov diffusion processes}

For our last application, we consider the two-dimensional transverse diffusion described by the SDE
\begin{equation}
    dX_t = -M X_t\,dt + \sigma\,dW_t,
\end{equation}
where $X_t \in \reals^2$, the noise matrix is proportional to the $2 \times 2$ identity matrix, i.e., $\sigma = \epsilon \mathbb{I}$, and the drift involves the matrix
\begin{equation}
    M = 
    \begin{bmatrix}
        \gamma & - \xi\\
        -\xi & \gamma
    \end{bmatrix}.
\end{equation}
The parameters $\epsilon > 0$, $\gamma > 0$, and $\xi \in \reals$ are set equal to $1$ for the application, in which case the process has a stationary current that circulates about the origin in a counterclockwise direction.

The observable under investigation is the nonequilibrium work, given by
\begin{equation}
    A_T = -\frac{1}{T} \int_{0}^{T} 2(D^{-1}M)^- X_t \circ dX_t,
\end{equation}
where we recall that $D = \sigma \sigma^\top$ is the covariance matrix and
\begin{equation}
    (D^{-1}M)^- = \frac{(D^{-1}M) - (D^{-1}M)^\top}{2}
\end{equation}
is the antisymmetric part of $D^{-1}M$. The rate function and SCGF for this process and observable have been found analytically (see \cite[Sec.~VB]{buisson2022b}), and the driven process is known to be another linear diffusion.

This information can be exploited in the representation of the control drift, but we instead showcase the use of the neural control drift to demonstrate the method's generality. Specifically, we use a fully connected feedforward neural network, which is a simple and general-purpose architecture, to represent the change in drift $\Delta \pF_\theta$. The parameterised control drift is thus
\begin{equation}
\pF_\theta(x) = -M x + \Delta \pF_\theta(x).
\end{equation} 
The network consists of two fully connected hidden layers of width 16, each followed by an activation function, and a final fully connected output layer that maps to a two-dimensional output. We use the SiLU activation function, which is smooth and continuously differentiable (unlike the more common ReLU activation function), so that the resulting control drift is smooth.

With this drift, trajectories of the control process are generated using the Euler--Maruyama method, which can easily be done in parallel to obtain independent trajectories. In principle, other SDE solvers can also be used, resorting to the adjoint method for computing the gradient if the computational graph becomes too large to use stochastic automatic differentiation. The TorchSDE library \cite{li2020}, building on the PyTorch framework, implements various differentiable SDE solvers, and includes the option to have the gradient computed via the adjoint method.

For our own simulations, we used the Euler--Maruyama integrator and automatic differentiation, obtaining the SCGF on a grid of $k$ values defined by $k_{\min} = -1.15$, $k_{\max} = 0.15$, and  $\Delta k = 0.05$. For the optimisation, we used the Adagrad optimiser with initial learning rate $\alpha_0 = 0.015$ over $N = 50$ epochs, simulating for each $n = 2\,000$ independent parallel trajectories up to the final time $T = 15$, using $\Delta t = 0.01$ for the integration time step. The final $M = 35$ epochs' estimates were then averaged to form the final estimates of the SCGF and rate function. 

\begin{figure*}[t]
    \centering
    \includegraphics[width=0.875\hsize]{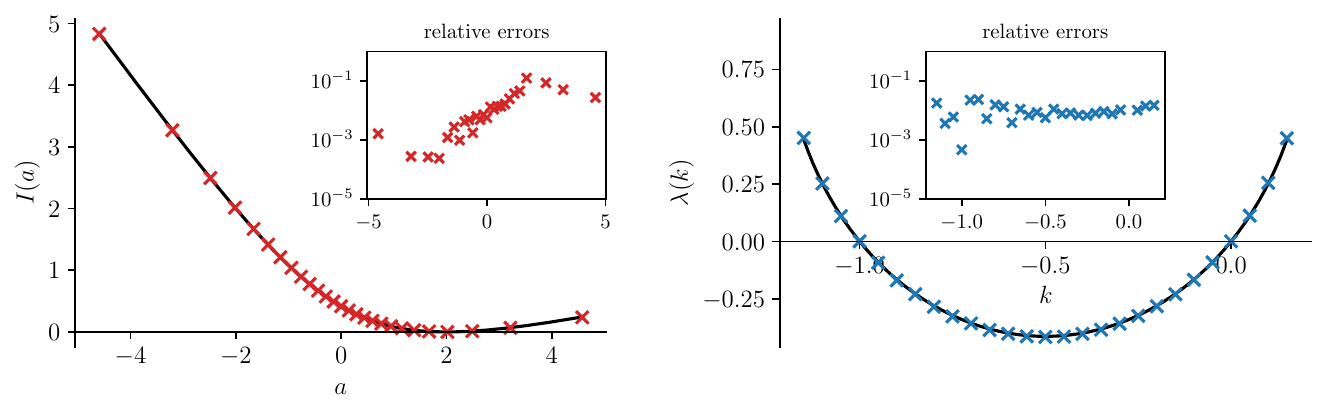}
    \caption{Rate function (left) and SCGF (right) for the nonequilibrium work of a two-dimensional transverse diffusion with parameters $\gamma = \xi = \sigma = 1$. Crosses denote the corresponding \ldstop estimates, and the insets show the relative errors.}
    \label{fig:SDEworkResults}
\end{figure*}

Figure~\ref{fig:SDEworkResults} compares the final results with the true large deviation functions obtained in \cite{buisson2022b}. The numerical estimates are very accurate, as can be seen, with relative errors between $10^{-1}$ and $10^{-4}$ (see inset). These are slightly larger than the relative errors in the discrete-time Markov chain example, which is expected, given that the space that must be sampled is now larger. In practice, the errors can be made smaller by increasing the simulation time, the number of trajectories simulated, as well as the number of epochs used to average the estimates.

\begin{figure*}[t]
    \centering
    \includegraphics[width=\textwidth]{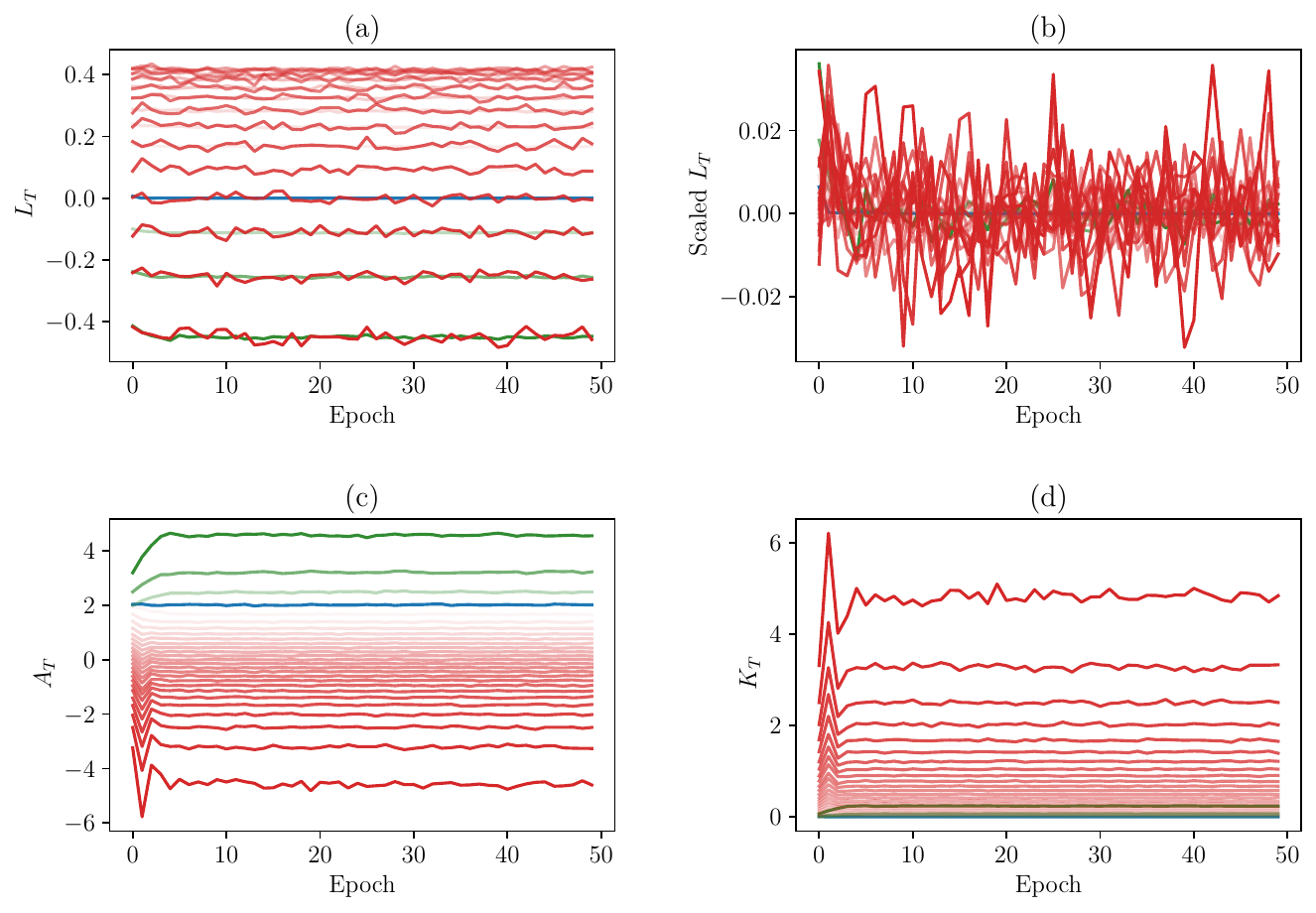}
    \caption{Convergence of \ldstop for the nonequilibrium work of a two-dimensional transverse diffusion with parameters $\gamma = \xi = \sigma = 1$. The colour scheme used is the same as in Fig.~\ref{fig:IIDexpConvergence}.}
    \label{fig:SDEworkConvergence}
\end{figure*}

The latter averaging is important. In Fig.~\ref{fig:SDEworkConvergence}(a)-(b), we see that the loss and scaled loss have on their own a fairly large variance even though they lead to accurate results for the large deviation functions. In fact, from the loss, it is difficult to see that the simulations have converged at all. In many cases, a better indicator of that convergence is provided by the traces of $A_T$ and $K_T$, which show in our case that convergence is reached around $10$ epochs; see Fig.~\ref{fig:SDEworkConvergence}(c)-(d). In general, we advise tracking the estimates of $A_T$ and $K_T$, in addition to the scaled loss, to determine whether convergence is reached. 

\section{Concluding remarks}
\label{secconc}

The examples that we have covered provide a template for applying the \ldstop method to other processes and observables. In future work, we will look at scaling the method up for estimating the large deviation functions of systems involving many particles, such as interacting particle systems evolving on lattices, used to describe transport processes, or interacting diffusions used for modelling chemical and biological systems driven in steady states by non-conservative or active forces (see \cite{yan2022} for an initial test case). 

By way of conclusion, we list and discuss next a few technical points about the method and how it differs from other methods, such as cloning, importance sampling, and APM.
\begin{itemize}
\item The main advantage of \ldstop is that it is based on independent trajectory simulations that do not need to be saved or kept in memory after each epoch, which makes it efficient, stable, and easy to implement. In cloning, a large population of coupled trajectories needs to be stored and managed globally, requiring extensive memory. Moreover, the managed trajectories become correlated over time, limiting the way error bars can be calculated (see the last point below). 

\item APM is also based on independent trajectory simulations, but represents the process simulated indirectly via an estimated eigenvector or eigenfunction. This limits its stability, especially for diffusions, which require that we calculate the derivative of an estimated function, a calculation step that amplifies the error present in that estimate \cite{ferre2018}.

\item From the test cases considered so far, we find that \ldstop does not suffer from slowing-down or sticking effects near dynamical phase transitions. This was observed for the random walk on the graph example (see Sec.~\ref{secMCld1}), as well as in the original proposal \cite{yan2022} of the method for an SDE having such a transition in the low-noise limit. The cloning method, by contrast, is known to require exponentially large populations to perform well near dynamical phase transitions. This can be mitigated by incorporating importance sampling in cloning \cite{nemoto2016}, something that \ldstop does directly without the need to manage a population of clones.

\item The \ldstop method is a principled importance sampling method that progressively learns or builds the optimal sampler (i.e., the driven process) in a direct way by minimising a computable loss in process space. Unlike naive or direct importance sampling, it does not require the construction of histograms in the double limit of large sample sizes and long integration times \cite{touchette2011}. The SCGF and rate function are obtained  by minimising a loss, and the minimal loss gives these functions directly, assuming again that this is indeed the global minimum.

\item The \ldstop method can be implemented using many established packages for representing complex functions (in particular, neural networks) and for performing large-scale optimisation, coming from recent research in machine learning. This makes the method easy to code and to deploy on dedicated CPU or GPU machines: the basic blocks for representing processes are available in packages such as PyTorch or TensorFlow, and so are the routines needed for performing the optimisation. In that, it follows many other methods that have been proposed recently for solving high-dimensional partial differential equations \cite{e2018,han2018,sirignano2018,cohen2023}, sampling many-particle systems \cite{wang2022,gabrie2022,mehdi2024}, controlling nonequilibrium systems \cite{engel2023}, and determining rare transition paths \cite{grafke2019,kikuchi2020,rotskoff2022,simonnet2023,grigorio2026}, among many other applications. The  idea common to these methods is to represent the problem to solve as an optimisation problem involving many parameters and, from there, to learn or fit the parameters using gradient descent or ascent methods. 

\item One of the limitations of \ldstop is that it has no convergence guarantees, in the same way that supervised learning based on neural networks is not guaranteed to converge. However, just as for neural networks, there are strong indications, based on simulations, that it should give accurate results provided that the representation used for simulating the controlled process has enough parameters to reach the ``theoretical'' optimal driven process. If the representation chosen is not rich enough for that purpose, then the estimates obtained provide bounds on the theoretical SCGF and rate function, as discussed. With this in mind, we suggest performing simulations with different representations so as to either confirm that the results are constant across representations or that one representation gives better results (viz., bounds). This can be done in parallel or sequentially by refining representations, that is, by adding more parameters. With the same idea, it is also possible to compute ``loose'' bounds on the large deviation functions with little computational cost, using simplified representations, such as linear drifts for diffusions \cite{nyawo2026}.

\item A final advantage of \ldstop, coming from the independent trajectories underlying this method, is that error bars can be calculated directly, either from the collection of $n$ trajectories simulated in parallel or from a single trajectory using batch mean methods \cite{asmussen2007}. We have not provided error bars in the present study because they are too small to show and exact results are available to compare with our numerical results. In a future study, we will provide more details on this important point. In the cloning algorithm, error bars are difficult to obtain because the population of trajectories kept in memory is correlated, and become more so as the algorithm advances. 
\end{itemize}

\section*{Acknowledgments}

H.T. thanks J. Yan and G. M. Rotskoff for the initial discussions and collaboration that led to this paper. The work of D.W.H.C. was funded by the Postgraduate Scholarship Programme (PSP) of Stellenbosch University.

\appendix
\section{Score gradient estimator derivation}
\label{secappend}

We derive in this appendix the score estimator of the gradient of the loss associated with the large deviations of sample means of independent random variables and Markov chains evolving in discrete time. The derivations are rather standard, but are useful for understanding the difference between this gradient estimator and the one obtained by automatic differentiation.

\subsection{Sample means of independent random variables}

We consider the sample mean $A_T$ in \eqref{eqsamplemean1} involving $T$ real random variables $(X_t)_{t\in [1:T]}$ assumed to be mutually independent and identically distributed with some density $p$. The KL divergence for this observable is shown in \eqref{eqKLiid1}. Parameterising the driven density as $\pp_\theta$, we are then faced with minimising the following expected loss:
\begin{equation}
\expec[\pp_\theta]{\ln \frac{\pp_\theta(X)}{p(X)}-kX}= \int_\reals \pp_\theta(x) \left[\ln\frac{\pp_\theta(x)}{p(x)}-kx\right].
\end{equation}

Taking the gradient of this expectation gives
\begin{equation}
\nabla_\theta \expec[\pp_\theta]{\ln \frac{\pp_\theta(X)}{p(X)}-kX} = \sum_{x} \pp_\theta(x) \nabla_\theta \ln \pp_\theta(x) \left(\ln \frac{\pp_\theta(x)}{p(x)} - k x\right)  + \sum_{x}\pp_\theta(x) \nabla_\theta \ln \pp_\theta(x)
    \label{eqscoreIIDexpec}
\end{equation}
It is easy to see that the second term on the right-hand side vanishes, so we are left with
\begin{equation}
\nabla_\theta \expec[\pp_\theta]{\ln \frac{\pp_\theta(X)}{p(X)}-kX} = \expec[\pp_\theta]{\left(\ln \frac{\pp_\theta(X)}{p(X)} - k X\right)\nabla_\theta \ln \pp_\theta(X) }.
\end{equation}

To estimate this gradient, we can generate from $\pp_\theta$ a sample $(\pX^{(i)})_{i\in [1:n]}$ of $n$ realisations of $X$ and compute
\begin{equation}
    \label{eqscoreIIDnobase}
    \nabla_\theta L_n = \frac{1}{n} \sum_{i = 1}^{n} L_i \nabla_\theta \ln \pp_\theta(X^{(i)}),
\end{equation}
where
\begin{equation}
    \label{eqscoreIIDlossi}
    L_i = \ln \frac{\pp_\theta(X^{(i)})}{p(X^{(i)})}-kX^{(i)}.
\end{equation}
Naturally, since the sample mean $A_T$ already contains $T$ such random variables, we can simply re-use them, replacing $X^{(i)}$ above with $X_t$, assuming $n=T$.

Note the difference between this score gradient estimator and the estimator obtained by taking the automatic derivative of the estimator of the loss $L_T = K_T-kA_T$, with $K_T$ given as in \eqref{eq:IIDKL}. In the score estimator, the gradient is only applied to the driven log-density. By contrast, the automatic derivative of $L_T$ leads to the calculation of that gradient, in addition to the gradient of the random variates generated from $\pp_\theta$, from which $L_T$ is calculated. The two approaches give the same result in expectation. The advantage of the score approach is that it can be applied to random variables whose generation is not differentiable in $\theta$, although $\pp_\theta$ itself must be differentiable. In practice, one should ensure the gradient does not accidentally pass through \eqref{eqscoreIIDlossi} or the counterpart for Markov chains. This can be accomplished in PyTorch through the \texttt{detach} method, which disconnects the variates, represented as tensors, from the computational graph.

As mentioned in Sec.~\ref{secgradient}, the score estimator often has a high variance, which can be mitigated by subtracting a baseline from the loss $L_i$ without changing its expectation; see \cite{sutton2018}. Here, we choose to remove as a baseline the expected loss coming from previous epochs in the minimisation, leading to the gradient estimator shown in \eqref{eqscoreIID}. 

\subsection{Discrete-time Markov chains}

The derivation of the score gradient for discrete-time Markov chains proceeds similarly to above. The main difference is that we start directly at the path level by considering the minimisation of the loss $\expec[\pmQ_T]{L_T}$ for a large but finite time $T$, where $L_T = K_T-k A_T$ and 
\begin{equation}
K_T = \frac{1}{T}\ln \frac{\pmQ_{\theta,T}[X]}{\pmP_T[X]}.
\end{equation}
This loss comes from the variational representation of $\lambda(k)$ shown in \eqref{eqvar2}. For the controlled Markov chain, the path distribution is written in terms of a parameterised transition matrix $\pPi_\theta$ and some initial distribution $P_0$ as
\begin{equation}
    \pmQ_{\theta,T}[X] = P_0(X_0) \prod_{t = 1}^{T} \pPi_\theta(X_{t-1}, X_t).
\end{equation}
This gives with a similar form for $\pmP_T[X]$ the expression of $K_T$ shown in \eqref{eqKTMC1}. Moreover, taking the logarithm followed by the gradient gives
\begin{equation}
    \label{eqpathscore}
    \nabla_\theta \ln\pmQ_{\theta,T}[X] = \sum_{t = 1}^{T} \nabla_\theta \ln \pPi_\theta(X_{t-1}, X_t).
\end{equation}

With these results, we write the gradient of the expected loss as
\begin{align}
\nabla_\theta \expec[\pmQ_{\theta,T}]{L_T} 
&= \nabla_\theta \sum_{x} \pmQ_{\theta,T}[x]\left[\frac{1}{T}\ln \frac{\pmQ_{\theta,T}[x]}{\pmP_T[x]}-kA_T\right] \nonumber\\
&=\sum_{x} \pmQ_{\theta,T}[x] L_T \nabla_\theta \ln \pmQ_{\theta,T}[x] 
+ \frac{1}{T} \sum_{x} \pmQ_{\theta,T}[x] \nabla_\theta \ln\pmQ_{\theta,T}[x] \nonumber\\
&= \expec[\pmQ_T]{L_T\sum_{t=1}^T \nabla_\theta \ln\pPi_\theta(X_{t-1},X_t)} + \expec[\pmQ_T]{\frac{1}{T} \nabla_\theta \ln\pmQ_{\theta,T}[X]},
\end{align}
the sums being on the paths of the Markov chains. As for sample means of independent random variables, the second term on the right-hand side above vanishes, so we obtain in the end
\begin{equation}
\nabla_\theta \expec[\pmQ_{\theta,T}]{L_T}  =\expec[\pmQ_T]{L_T\sum_{t=1}^T \nabla_\theta \ln\pPi_\theta(X_{t-1},X_t)}
\end{equation}
for the theoretical gradient of the loss.

The empirical estimation of this gradient follows by noting that $L_T$ concentrates in the long-time limit to a constant, as the driven process is assumed to be ergodic. Therefore, our natural estimator of the gradient is 
\begin{equation}
\nabla_\theta L_T = L_T\sum_{t=1}^T \nabla_\theta \ln\pPi_\theta(\pX_{t-1},\pX_t).
\label{eqRest1}
\end{equation}
This is the REINFORCE gradient used in reinforcement learning when dealing with policy gradient methods for Markov decision processes \cite{williams1992,sutton2018}. As already noted, the variance of this estimator is reduced in practice by removing a baseline, which we choose here to be an average loss $\bar{L}$ taken over a certain number of previous epochs, leading to the estimator shown in \eqref{eqscoreMCaveloss}.

In practice, the variance can be further reduced without biasing the estimator by subtracting components of the loss in \eqref{eqRest1} that come before the state $X_{t-1}$ inside the sum. This leads us to consider a new estimator having the form
\begin{equation}
\nabla_\theta L_T = \sum_{t=1}^T L_{t:T}\nabla_\theta \ln\pPi_\theta(\pX_{t-1},\pX_t),
\end{equation}
where $L_{t:T}$ is the partial loss estimated from the times $t$ to $T$ rather than from $t=0$ to $T$. To avoid considering partial losses that are too short to be self-averaging, we also truncate the sum above to the first half of the terms. Subtracting yet again an epoch-based baseline then gives
\begin{equation}
\nabla_\theta L_T = \sum_{t=1}^{\lfloor T/2 \rfloor} (L_{t:T}-\bar{L})\nabla_\theta \ln\pPi_\theta(\pX_{t-1},\pX_t),
\label{eqscore}
\end{equation}
where $\lfloor T/2 \rfloor$ is the integer part of $T/2$. This truncated gradient estimator, which recovers theoretically the true gradient in the long-time limit, is compared with the one obtained from the full sum up to time $T$ for an example of jump processes in Sec.~\ref{secJP1}.

\bibliography{StochOptim}

\end{document}